\documentclass[aps,prl,twocolumn,showpacs]{revtex4}

\begin{document}

\title{Accelerated expansion of the universe driven by  tachyonic matter}

\author{T.Padmanabhan}
\email[]{ nabhan@iucaa.ernet.in }
\homepage[]{http://www.iucaa.ernet.in/~paddy}
\affiliation{IUCAA, Ganeshkhind, Pune, India 411 007}


\date{\today}

\begin{abstract}
It is an accepted practice in cosmology to invoke a scalar field with potential $V(\phi)$
when observed evolution of the universe cannot be reconciled with theoretical
prejudices. Since one function-degree-of-freedom in the expansion factor $a(t)$
can be traded off for the function $V(\phi)$, it is {\it always} possible
to find a scalar field potential which will reproduce a given evolution. 
I provide a recipe for determining $V(\phi)$ from $a(t)$ in two cases:
(i) Normal scalar field with Lagrangian  ${\cal L} = (1/2)\partial_a\phi \partial^a\phi - V(\phi)$ used
in quintessence/dark energy models;
(ii) A tachyonic field with Lagrangian  ${\cal L} = -V(\phi) [ 1- \partial_a\phi \partial^a\phi]^{1/2} $,
motivated by recent string theoretic results. In the latter case, it is possible
to have accelerated expansion of the universe during the late phase in certain 
cases. This suggests a   string theory based interpretation of the current phase of the 
universe with tachyonic condensate acting as effective cosmological constant. 

\end{abstract}


\maketitle

  \section{1. Introduction }
   \noindent
    The stress tensor $T^a_b$ for any source term in a Friedmann universe, 
    described by an expansion factor $a(t)$, must have the form $T^a_b (t) = {\rm dia}\ [\rho(t),-p(t),
    -p(t),-p(t)]$. Given an equation of state which specifies $p$ as a function of $\rho$,
    we will be left with two degrees of freedom $a(t)$ and $\rho(t)$ which can be determined
    by two independent Einstein's equations for the Friedmann model. The situation, however,
    is slightly different if the source is described by an ``adjustable function'' as in the case of 
    scalar fields. If, for example, the source is described by a scalar field with a Lagrangian
    ${\cal L} = (1/2)\partial_a\phi \partial^a\phi - V(\phi)$, then it is possible to choose
    $V(\phi)$ in order to have a specific evolution for the universe.  Given any $a(t)$ it is
     {\it always} possible to obtain a $V(\phi)$ such that it 
    results in a consistent
    cosmological evolution. In fact, this can be achieved even in the presence of 
    other energy densities  in the universe (like matter, radiation etc.) in addition to the 
    scalar field. This should not be surprising, since the existence of a free function $V(\phi)$
    allows a trade off with another function $a(t)$. 
    
    Recently, it has been suggested that \cite{asen} the tachyonic condensate in a class of string
    theories can be described by an effective scalar field with a Lagrangian of the form
    ${\cal L} = -V(\phi) [ 1- \partial_a\phi \partial^a\phi]^{1/2} $. The evolution of this 
    condensate can have cosmological significance which may be worth exploring \cite{gibb}.
    Since this Lagrangian also has a potential function $V(\phi)$, it seems reasonable to 
    expect that {\it any} form of cosmological evolution (that is, any $a(t)$) can be obtained
    with the tachyonic field as the source by choosing $V(\phi)$ ``suitably''.
    It turns out that this is indeed true. 
    
   I will outline a recipe for constructing $V(\phi)$ given a particular form of evolution for the 
    universe $a(t)$ in the two cases (normal scalar field and tachyonic field) mentioned above.
    The first case corresponds to quintessence/dark energy models and has been a 
    favourite pastime of the cosmologists in the last several years [for a sample of references, see \cite{varun}]. In this case, there is very little
    (independent) constraint on $V(\phi)$ and hence it is not possible to evaluate 
    the relative merits of different choices for $V(\phi)$.  In the case of tachyonic scalar field, there 
    are some constraints on the form of $V(\phi)$, especially on the asymptotic behaviour,
    which could rule out certain class of cosmological models. It is nevertheless possible to construct  
    several interesting models  satisfying the asymptotic constraints on
    $V(\phi)$. In particular, it may be possible to have a rapidly accelerated phase of expansion
    for the universe at late times which seems to have some observational
    support. 
    
 \section {2. Recipe for the scalar field potential}
   
   \noindent
    Consider a $k=0$ universe with a normal scalar field having a potential $V(\phi)$ as the 
    source. 
    We assume that the evolution of the universe is already specified so that $a(t)$, $H(t) \equiv (\dot a/a)$ ...
    etc. are known functions of time and we need to determine $V(\phi)$ such that Friedmann
    equations
    \begin{equation}
     \left(\frac{\dot a}{a}\right)^2 = \frac{8\pi G}{3} \rho; \quad  \left(\frac{\ddot a}{a}\right) =- \frac{4\pi G}{3} 
     (\rho +3p)
     \end{equation}
     as well as the equation of motion for the scalar field 
     \begin{equation}
      \ddot \phi + 3H\dot \phi = - \frac{dV}{d\phi}
      \label{phiddot}
     \end{equation}
     are satisfied. (Of course, only two of these three equations are independent when the universe is
     driven by  a single source.)
    In a Friedmann universe, $\phi(t, {\bf x}) = \phi(t)$ and the energy density and pressure
    of the scalar field is given by 
  \begin{equation}
   \rho_\phi = \frac{1}{2} \dot \phi^2 + V(\phi); \qquad P_\phi = \frac{1}{2} \dot \phi^2 - V(\phi)
   \label{qrhop}
   \end{equation}
   It is convenient to define a time dependent parameter $w(t)$ by the relation 
$w(t) \equiv  P_\phi (t) / \rho_\phi (t) $.
The equation of motion for the scalar field, written in the form
$d(\rho a^3) = - w \rho d(a^3)$ can be integrated 
   to give 
   $ (\dot \rho_\phi/\rho_\phi)= - 3 H (1+w)$.
   The Friedmann equation, on the  other hand, 
   gives
    $\rho_\phi \propto H^2$ so that 
$(\dot \rho_\phi/\rho_\phi) = 2 (\dot H/H)$. Combining the two relations
we get
\begin{equation}
 1+ w(t) = - \frac{2}{3} \frac{\dot H}{H^2}
\label{qopw}
\end{equation}
thereby determining $w(t)$.
(Note that we have not used the specific form of the source so far; so, this equation will be satisfied by
any source in a FRW model.)
From the definition of $w$ and (\ref{qrhop}), it follows that $\dot \phi^2/ 2V = (1+w) (1-w)^{-1}\equiv f(t)$, say. Writing this as $\dot \phi^2 = 2fV$, differentiating
with respect to time and using (\ref{phiddot})
 we find that
 \begin{equation}
 \frac{\dot V}{V} = - \frac{\dot f + 6 Hf}{1+f}
 \end{equation}
Integrating this equation and using the definition of $f(t)$ and
equation (\ref{qopw}) we get
\begin{equation}
 V(t) = \frac{3H^2}{8\pi G} \left[ 1+ \frac{\dot H}{3H^2}\right]
\label{qmainv}
\end{equation}
Substituting back in the relation $\dot \phi^2 = 2fV$, we can
determine $\phi(t) $ to be 
\begin{equation}
 \phi(t) = \int dt \left[ -\frac{\dot H}{4\pi G}\right]^{1/2}
 \label{qmainphi}
 \end{equation}
Equations (\ref{qmainv}) and (\ref{qmainphi}) completely solve the problem
of finding a potential $V(\phi)$ which will lead to a given
$a(t)$. These equations determine $\phi(t)$ and $V(t)$ in terms of $a(t)$
thereby implicitly determining $V(\phi)$.

In fact, the same method works even when matter other than scalar field
with some {\it known} energy density $\rho_{\rm known} (t)$ present in the universe.
In this case, equations (\ref{qmainv}) and (\ref{qmainphi}) generalizes to
\begin{equation}
 V(t) = \frac{1}{16\pi G} H (1-Q)\left[6H + \frac{2\dot H}{H} - \frac{\dot Q}{1-Q}\right]
 \end{equation}
 \begin{equation}
 \phi (t) = \int dt \left[ \frac{H(1-Q)}{8\pi G}\right]^{1/2} \left[\frac{\dot Q}{1-Q} - \frac{2\dot H}{H}\right]^{1/2}
 \end{equation}
where $Q (t) \equiv [8\pi G \rho_{\rm known}(t) / 3H^2(t)]$.

As an example of using (\ref{qmainv}) and (\ref{qmainphi}), let us consider
a universe in which $a(t) = a_0 t^n$. Elementary algebra now gives
the potential to be of the form 
\begin{equation}
 V(\phi) = V_0 \exp \left( - \sqrt{ \frac{2}{n}} \, \frac{\phi}{M_{\rm Pl}}\right)
 \label{vfortn}
 \end{equation}
where $V_0$ and $n$ are constants and $M_{\rm Pl}^2=1/8\pi G$.
The corresponding evolution of $\phi (t)$ is given by
\begin{equation}
 \qquad \frac{\phi(t)}{M_{\rm Pl}} = \sqrt{2n} \ln \left( \sqrt {\frac{V_0}{n(3n-1)}}  \, \frac{t}{M_{\rm Pl}}\right)
 \end{equation}
As a second example, 
consider an evolution of the form
\begin{equation}
 a(t) \propto \exp(\alpha t^f),\quad f= \frac{\beta}{4+\beta}, \quad 0< f < 1, \quad \alpha>0
 \label{extwo}
 \end{equation}
In this case, we can determine the potential to be  
\begin{equation}
 V(\phi) \propto \left( \frac{\phi}{M_{\rm Pl}}\right)^{-\beta} \left( 1 - \frac{\beta^2}{6} \, \frac{M_{\rm Pl}^2}{\phi^2}\right)
 \end{equation}
where $\beta$ is a constant. The two potentials described above have been used extensively
in inflationary models. 

In fact, virtually all other potentials used in quintessence/dark energy models
for the universe can be obtained by the recipe given above. Since one  seldom worries seriously
about the microscopic origin of $V(\phi)$ in these models, it may be mathematically more convenient to choose
one's favourite cosmological evolution in terms of $a(t)$ and then construct $V(\phi)$ and study its properties.

  \section{3. Recipe for tachyonic potential}
  
  \noindent
  Consider next a universe with a given $a(t)$ and a tachyonic source with the Lagrangian
  ${\cal L} = -V(\phi) [ 1- \partial_a\phi \partial^a\phi]^{1/2} $. When $\phi =\phi(t)$, the energy
  density and pressure are given by
  \begin{equation}
  \rho = {V(\phi) \over \sqrt{1-\dot\phi^2}}; \quad p = -V \sqrt{1-\dot\phi^2}
  \label{rhoandp}
  \end{equation}
  In this case, the ``reverse engineering'' to determine $V(\phi)$ from $a(t)$ is almost trivial.
   For {\it any}
  source with a parameter $w(t)$, we must have 
  \begin{equation}
  {\dot \rho\over \rho} = - 3 H(t) (1+w) = {2 \dot H\over H}
  \end{equation}
  leading to (\ref{qopw}).   On the other hand,  for the tachyonic model,  $p(t)/\rho(t) \equiv w(t) =  \dot \phi^2-1$. Combining these, we can determine $\dot\phi^2=-(2/3)(\dot H/H^2)=(2/3)(dH^{-1}/dt)$ in terms of $H$ 
  and obtain 
  \begin{equation}
  \phi(t) = \int dt \left( -{2\over 3}{\dot H\over H^2}\right)^{1/2}
  \label{finalone}
  \end{equation}
   Multiplying the two equations in (\ref{rhoandp}) and using (\ref{qopw}) and the Friedman equation, we get 
   \begin{equation}
   V=(-w)^{1/2} \rho = {3H^2 \over 8\pi G} \left( 1 + {2\over 3}{\dot H\over H^2}\right)^{1/2}
   \label{finaltwo}
   \end{equation}
   Equations (\ref{finalone}) and (\ref{finaltwo}) completely solve the problem. Given any 
   $a(t)$, these equations determine $V(t)$ and $\phi(t)$ and thus the potential $V(\phi)$. Equation (\ref{finalone}) also implies that $\dot H<0$ for these models.
  
   Note the  similarity between the forms of $V(\phi)$ in (\ref{finaltwo}) and (\ref{qmainv}).
   The fact that both tachyonic and normal scalar field potentials can be used to drive 
   the expansion of the universe, suggests that --- as far as cosmological evolution
   is concerned --- there exists a mapping between the two potentials directly.
  For example, we found that an exponentially decaying potential for the normal 
   scalar field leads to power law growth for $a(t)$. I will now show
   that a tachyonic potential of the form $V(\phi) \propto \phi^{-2}$ will
   lead to the same kind of cosmological evolution. 
   
   Consider  a universe with power law expansion
   $a= t^n$. In this case, $(\dot H/H^2)$ in equation (\ref{finalone}) is a constant
making $\dot \phi $  a constant. The complete solution
   is given by
   \begin{equation}
   \phi(t) = \left({2\over 3n}\right)^{1/2} t + \phi_0; \quad
   V(t) = {3n^2\over 8\pi G}\left( 1- {2\over 3n}\right)^{1/2} {1\over t^2}
   \end{equation}
   where $n>(2/3)$. [I will comment on the $n=(2/3)$ case later on.]
   Combining the two, we find the potential to be 
   \begin{equation}
    V(\phi) = {n\over 4\pi G}\left( 1- {2\over 3n}\right)^{1/2}
   (\phi - \phi_0)^{-2}
   \label{tachpot}
   \end{equation}
   For such a potential, it is possible to have arbitrarily rapid expansion with large $n$. [It is also
   possible to reproduce the evolution in (\ref{extwo}) by a more complicated choice of the potential,
   of the form $V\propto \phi^x(1+c_1\phi^y)^{1/2}$. The reverse-engineering procedure is exactly
   the same].]
   
    If $\phi_a\phi^a\ll 1$ the tachyonic Lagrangian can be approximated by the form
   \begin{equation}
   {\cal L} = -V(\phi) [ 1- \partial_a\phi \partial^a\phi]^{1/2}\approx {1\over 2}\psi_a\psi^a -U(\psi)
   \label{mapping}
   \end{equation}
   with 
   \begin{equation}
   \psi=\int \sqrt{V(\phi)}d\phi; \quad U(\psi)=V[\phi(\psi)]
   \end{equation}
   In our case, $V(\phi)=A/\phi^2$ (say), giving $\psi=\sqrt{A}\ln \phi$ and
   $U(\psi)\propto \exp(-2\psi/\sqrt{A})$. Curiously, this has the same {\it form} of
   the potential we found in (\ref{vfortn}) for a normal scalar field to produce the
   $a=t^n$ evolution though the coefficient of $\psi$ matches with
that in  (\ref{vfortn}) only when $n\gg 1$.  In this limit, the mapping in (\ref{mapping}) 
becomes increasingly accurate.  
   
   The potential in (\ref{tachpot}) has the reasonable behaviour of $V\to 0$ as 
   $\phi \to \infty$ though its form for small and intermediate values of $\phi$ is not 
   supported by string theory. It is however possible to show that the asymptotic
   form of the evolution is still given by the solutions found above. To see this,
   assume that the late time behaviour of $\phi$ is given by 
   \begin{equation}
   \phi(t) = \left({2\over 3n}\right)^{1/2}\, t + B e^{-Ct} = \left({2\over 3n}\right)^{1/2}\, t + {\cal O} (e^{-Ct})
   \end{equation}
   where $n, B, C$ are constants with $(2/3)\le n $. This implies that we take $\phi(t)$
   to grow proportional to $t$ asymptotically with exponentially small corrections.
   In this case, it is possible to repeat the above analysis and show that asymptotically
   (at late times), we have the following behaviour:
   \begin{equation}
   a(t) \approx t^{n} \exp\left[  {\cal O}({e^{-Ct}\over t})\right]
   \end{equation}
   and
   \begin{equation}
   V(\phi) \approx {n\over 4\pi G}\left( 1- {2\over 3n}\right)^{1/2} {1\over \phi^2}[1+{\cal O}({e^{-Ct}\over t})]
   \end{equation}
   The asymptotic form for $a(t)$ is essentially a power law
   found before with exponentially small corrections.
   The time scale for the validity of asymptotic solution is determined by
   $(1/C)$ which is a parameter that can be fixed independent of 
   $n$. This will allow one to ignore exponentially subdominant
   terms even when large values of $n$ are invoked.
   
   This analysis assumes that the asymptotic ``velocity'' of the scalar field
   is given by $\dot \phi (\infty) = (2/3n)^{1/2}$ if the expansion of the universe
   at late times is given by $a(t) = t^n$. 
   This, in turn, requires the tachyonic potential $V(\phi)$ to have the asymptotic
   form $A/\phi^2$ without any restriction on $A$. To have accelerated expansion, we must
   have reasonably large $n$ requiring $\dot \phi(\infty)$ to be sufficiently less than
   unity --- which is possible if $A$ is non zero and arbitrary. 
   
   The situation is different if there are  string theoretic reasons to expect   $\dot \phi(\infty) =1$ asymptotically.
   Since equation (\ref{finalone})  can be equivalently written as
   \begin{equation}
   H^{-1}(t) = {3\over 2} \int \dot \phi^2 dt
   \end{equation}
    $\dot \phi(\infty) =1$ asymptotically will imply $a(t)\propto t^{2/3}$. 
   Then we must have $n=(2/3)$  which is 
   just a dust dominated, zero pressure, expansion law. 
   The cosmological implications are very different depending on whether string theory requires
   the condition  $\dot \phi(\infty) =1$ asymptotically or not.  Let me discuss the different possibilities
   briefly.
   
   If it is possible to have
   $\dot \phi(\infty)\ne 1$ asymptotically, then an attractive scenario emerges
   in which the tachyonic condensate could be driving the acceleration of the universe
   and mimicking the cosmological constant.
   In this case, one is interpreting the asymptotic evolution to be applicable during the 
   current phase of the universe, in the range of redshifts, say, $0<z<3$. 
   In a realistic model, one may need to worry about the timescales in string theory
   (presumably $\simeq t_{\rm Planck}$) vis-a-vis the cosmological timescale. But this issue will
   arise in any attempt to use the tachyonic condensate in the late phases in the evolution of the
   universe, like --- for example --- as some kind of dark matter.
   
   An easy way out is to use the solutions in the early universe.
   The solutions
   I have obtained, of course, could also be used to provide a power law inflation
   in the very early universe if  $t\to \infty$ is interpreted 
   merely as $t\gtrsim  t_{\rm Planck}$. It may be easier to have non vanishing
   pressure for the tachyonic condensate for $t\gtrsim  t_{\rm Planck}$ than during the 
   current epochs of the universe.  (I think it would be nicer if string theory
   could provide an effective cosmological constant at the current epoch --- which has
   {\it some} observational support --- rather
   than merely provide yet another inflaton field. But the procedure outlined here can be
   used to construct a wide class of inflationary solutions as well.) 
   
   If, on the other hand, string theory demands $\dot \phi(\infty)= 1$ asymptotically, then
  $V\to 0$ asymptotically in such a manner as to give finite $\rho$ and zero pressure.
   The tachyonic condensate can then contribute, say, $\Omega =0.7$ in the universe
   and --- together with clustered normal matter contributing $\Omega =0.3$ ---
   can lead to a $\Omega=1$ universe. Such a model is an extreme form of mixed dark matter model
   with a very smoothly distributed component at large scales. It is possible that
   the model is consistent with CMBR and galaxy clustering data but
   will contradict the supernova data if it is interpreted as indicating $\ddot a>0$ in 
   the recent past. (Because $\ddot a <0$ when $n<1$,  this model will not be
   accelerating.) Given the observational uncertainties, it may still be worth
   studying the consequences of such a model.  
   
   Even in this case [with $\dot \phi(\infty)= 1$ asymptotically], it may be possible to provide an accelerated expansion for the 
   universe in the recent past if one considers more complicated potentials.
   For example, one can  arrange matters
such that $n\gg 1$ in the redshift range of $3>z>0$, say, with the asymptotic regime
  of $n=(2/3)$ coming into effect only 
 in the future. The procedure developed in this paper can be used to obtain
 suitable $V(\phi)$ which will ensure such a scenario.

   All these emphasize
   the need to understand 
   the constraints on $V(\phi)$ from string theory. 
   
   I thank Ashoke Sen for useful discussions.

\end{document}